\documentstyle[12pt]{report}
\begin{document}


\pagestyle{empty}

\renewcommand{\thefootnote}{\fnsymbol{footnote}}


\begin{flushright}
{\small
SLAC--PUB--7504\\
hep-th/\\
May 1997\\}
\end{flushright}

\begin{center}
{\bf\large   
Deformations of the SUSY SU(5) Theory with an Antisymmetric Tensor\footnote{Work supported by
Department of Energy contract  DE--AC03--76SF00515.}}

\smallskip

Chih-Lung Chou\\
Stanford Linear Accelerator Center\\
Stanford University, Stanford, California 94309\\
and\\
Applied Physics Department\\
Stanford University, Stanford, California 94309\\
alexchou@leland.stanford.edu\\
\medskip

\end{center}

\vfill

\begin{center}
{\bf\large   
Abstract }
\end{center}

\indent The N=1 supersymmetric gauge SU(5) theory with one antisymmetric tensor, $n+3$ fundamentals and $n+4$ antifundamentals has dual magnetic theories in the infrared.  By introducing extra singlet fields and tree level superpotential terms to the electric SU(5) theories, we are able to make the magnetic dual theories flow to the known $SU(n) \times SU(2)$ gauge theories which break supersymmetry dynamically.  In the $n=2$ case, the lifting of pseudo-flat direction is estimated qualitatively by using dual operator mappings.

\vfill\eject
\pagestyle{plain}
\pagenumbering{arabic}
{\LARGE{\noindent 1    Introduction}}
\vspace{0.5cm}

Some N=1 supersymmetric gauge theories in the infrared can have dual descriptions of different gauge groups and matter content[1].  For example, a supersymmetric gauge $SU(N_c)$ theory with $N_f$ $(N_f \ge N_c +2)$ number of flavors is physically indistinguishable in the infrared from that of gauge group $SU(N_f -N_c)$ with $N_f$ flavors and $SU(N_c)$ bound states. Generalized electric-magnetic duality has also been found to exist in those supersymmetric gauge theories with higher-rank tensor multiplets[2].  The original electric theories are deconfined to expanded electric theories with product gauge groups such that one of the product groups has a dual description in the infrared.  These higher-rank tensor multiplets are viewed as compound states of strongly coupled gauge groups. Tree level superpotential terms necessarily appear in the dual theory to eliminate extra states in the dual moduli space. For those theories in the non-Abelian Coulomb phase, the stronger the electric gauge coupling is, the weaker the dual magnetic gauge coupling. These phenomena are supersymmetric versions of the generalized electric-magnetic duality in the supersymmetric non-Abelian gauge theories. 

Supersymmetry breaking is also shown to have dual descriptions in the N=1 supersymmetric gauge theories[3]. Recent developments on dynamical supersymmetry breaking (DSB) gauge models[4] show that non-perturbative superpotential terms or quantum constraints on moduli space generated by gauge dynamics may result in supersymmetry breaking.  Supersymmetry breaking phenomena are thus highly non-trivial phenomena in these DSB models.  However, the same supersymmetry breaking phenomena could possibly be as trivial as those of O'Raifeartaigh models when their dual magnetic descriptions in the infrared are examined. Non-trivial phenomena thus can become trivial through duality and confining. Generically, duality and confining mechanism can interrelate some of the DSB gauge models as various descriptions in different limits[5].      

The SUSY SU(2k+1) models with an antisymmetric tensor have been shown by Pouliot[2] to have dual descriptions in the infrared once the suitable tree level superpotential is added in. It has also been shown that by giving an expectation value of rank one to the matrix of the electric quark-antiquark condensate, the magnetic dual theory can flow to the first of Berkooz' duals[2]. In this paper, we investigate only the supersymmetric gauge SU(5) models with matter content of an antisymmetric tensor,$n+3$ fundamentals and $n+4$ antifundamentals. Tree level superpotential terms are given to make the theories break supersymmetry classically.  Although we cannot really analyze the theories in their strongly coupled regime, some phenomena can still be seen by looking up their dual magnetic theories.  In section 2 we discuss the $n \ge 3$ cases.  The dual magnetic theories are shown to flow to the DSB models with SU($n$)$\times$SU(2) product gauge groups after integrating out heavy matter multiplets. In section 3, the n=2 case is investigated. The dual magnetic theory is shown to be the supersymmetric gauge SU(2)$\times$SU(2) model which breaks supersymmetry dynamically.  The Yukawa coupling superpotential terms necessarily appear in the dual magnetic theory with classically pseudo-flat direction and large Yukawa coupling constants. The pseudo-flat directions are expected to be lifted by quantum corrections[6] which is similar to the Coleman-Weinberg mechanism[7] of corrections to the K\"{a}hler potential. Large Yukawa couplings and gauge couplings prevent further perturbative analysis and thus tell us nothing about the magnitude of the quantum lifting. However, if we make the naive estimation of the lifting of the flat direction through operator mapping between the original electric theory and the final dual magnetic theory, we find that the scale of the lifting of the pseudo-flat direction $\lambda {\bar a}$ is much less than the gauge dynamics scale $\Lambda_{32}$ and thus approaches to the origin.  In section 4, we give our conclusions.   

\vspace{0.5cm}
{\Large {\noindent 2. Case 1: $n \geq 3$ }}

\vspace{0.4cm}

It is known that the supersymmetric SU(5) gauge theory with one antisymmetric tensor A, $n+3$ fundamentals $Q_i$ and $n+4$ antifundamentals $\bar Q_a$ has a magnetic dual theory in the far infrared, for $n \ge 2$. With suitable tree level superpotential terms, this theory has dual descriptions and its supersymmetric moduli space can be determined in various limits. If the theory is modified by adding in extra singlet fields and additional tree level superpotential terms, then by adjusting parameters in the superpotential we can have the theory in the Higgs phase and find its ground states in the semi-classical limit.  For some cases, it is not hard to add in a tree level superpotential that leads to broken supersymmetry in the weak coupling limit. On the other hand, if the parameters are varied continuously and lead the theory to its strong coupling limit, the electric theory will not be suitable for quantitative analysis ,due to the non-perturbativeness. Fortunately, the theory that we consider has dual magnetic descriptions which may be weakly coupled while the electric SU(5) gauge group gets strong.  Some analytic work thus is possible by using dual magnetic descriptions. Although we cannot analyze the electric theory directly in the strong coupling limit, we still expect supersymmetry to be broken in this limit since supersymmetry is broken classically.

For $n<10$, the SU(5) coupling is asymptotically free with dynamical scale $\Lambda_5$. If the superpotential is perturbed by

\vspace{0.2cm}

$W=S_{n+3, n+4}(Q_{n+3}{\bar Q_{n+4}} - m^2)$,  $m \not= 0$, \hfill (1)

\vspace{0.2cm}

\noindent then the electric theory will be higgsed to SU(4) and its low energy magnetic dual is the first of Berkooz' dual once the additional superpotential term proportional to $AAQ_i$ is added in.  In this paper, instead of adding in Eq.(1), we choose the following three level superpotential

\vspace{0.2cm}
 
$W=S_{ia}(Q_i{\bar Q_a}-m^2\delta_{ia})$, $m \not= 0$,  \hfill (2)

\vspace{0.2cm}

\noindent where $i = 1, \cdots , n+3$ and $a = 1, \cdots , n+4$. $S_{ia}$ are the extra singlets of SU(5) and $m$ is the mass parameter input by hand. When the parameter $m$ is assumed much larger than $\Lambda_5$, the theory is in the weakly coupled regime and the elementary fields can be treated as suitable degrees of freedom of the SU(5) dynamics. When $n \ge 3$, the equation ${{\partial W} \over {\partial S_{ia}}}=0$ cannot have a solution in consistent with supersymmetry in this semi-classical limit and therefore supersymmetry is broken in this limit.  However, when $m$ is varied continuously and approaches the strongly coupled regime, the supersymmetry breaking phenomenon does not necessarily hold since the electric theory becomes non-perturbative and the semi-classical analysis fails in this limit. 
	
\indent For some supersymmetric gauge theories in the strongly coupled regime, the dual magnetic theories may be the suitable ones to analyze.  The deconfining method[8] is widely used to find out the magnetic dual theories of those electric ones with anti-symmetric (symmetric) tensors. The antisymmetric tensor A can be thought of as a bound state $y \cdot y$ due to the strong SU(2) gauge group. The SU(2) scale $\Lambda_2$ is thus assumed to be much larger than $\Lambda_5$. The transformation properties of the matter content  of the deconfined electric theory are listed below.

\bigskip
\begin{tabbing}
\hspace{3cm}		  \=SU(5) \hspace{1cm} 		\= SU(2) \hspace{1cm} 	\= SU($n+3$) \hspace{1cm} 	\=SU($n+4$) \\
\hspace{1cm} $y$\hspace{1cm}\> 5 \hspace{1cm} 	   	\>2  \hspace{1cm}     	\>1   \hspace{1cm}      	\>1       \\
\hspace{1cm} z\hspace{1cm}\>1   \hspace{1cm}    	\> 2   \hspace{1cm}    	\> 1   \hspace{1cm}    		\>1 	 \\
\hspace{1cm}${\bar p}$\hspace{1cm} \>${\bar 5}$ \hspace{1cm}\>1 \hspace{1cm}  	\> 1 \hspace{1cm}    		\>1	 \\
\hspace{1cm} Q\hspace{1cm}\> 5	\hspace{1cm}		\> 1 \hspace{1cm}	\>$n+3$	\hspace{1cm}		\>1	\\
\hspace{1cm}${\bar Q}$\hspace{1cm}\>${\bar 5}$\hspace{1cm}	\>1\hspace{1cm}	\>1     \hspace{1cm}		\>$n+4$    
\end{tabbing}

\bigskip

\vspace{0.2cm}

\noindent The expanded theory has gauge groups $SU(5) \times SU(2)$ and flavor groups $SU(n+3) \times SU(n+4)$ with an additional tree level superpotential term ${\bar p}yz$.  When $n \geq 2$, there are $n+5$ flavors of the SU(5) gauge group and thus the theory at the origin of the moduli space has dual descriptions.  The dual magnetic theory has product gauge groups $SU(n)\times SU(2)$ with matter fields which transform as:

\begin{tabbing}
\hspace{3cm}		  \=SU($n$) \hspace{1cm} 		\= $SU(2)_1$ \hspace{1cm} 	\= SU($n+3$) \hspace{1cm} 	\=SU($n+4$) \\
\hspace{1cm} ${\bar x}$\hspace{1cm}\> ${\bar n}$ \hspace{1cm} 	   	\>2  \hspace{1cm}     	\>1   \hspace{1cm}      	\>1       \\
\hspace{1cm} ${\bar q}$\hspace{1cm}\>${\bar n}$   \hspace{1cm}    	\> 1   \hspace{1cm}    	\> ${\overline{n+3}}$   \hspace{1cm}    		\>1 	 \\
\hspace{1cm} p\hspace{1cm} \>n	 \hspace{1cm}	\>1 \hspace{1cm}  	\> 1 \hspace{1cm}    		\>1	 \\
\hspace{1cm} q\hspace{1cm}\> n	\hspace{1cm}		\> 1 \hspace{1cm}	\>1	\hspace{1cm}		\>${\overline{n+4}}$	\\
\hspace{1cm}${\bar l}$\hspace{1cm}\>1\hspace{1cm}	\>2\hspace{1cm}	\>1     \hspace{1cm}		\>$n+4$ \\
\hspace{1cm}M\hspace{1cm}	\>1\hspace{1cm}	\>1\hspace{1cm}	\>$n+3$ \hspace{1cm}	\>$n+4$ 	\\
\hspace{1cm}$B_1$ \hspace{1cm}	\>1 \hspace{1cm}	\>1 \hspace{1cm}	\>$n+3$ \hspace{1cm}	\>1	\\
\hspace{1cm}S\hspace{1cm}	\>1 \hspace{1cm}	\>2 \hspace{1cm}	\>1   \hspace{1cm}      \>1    \\
\hspace{1cm}Z \hspace{1cm}	\>1 \hspace{1cm} 	\>2 \hspace{1cm} 	\>1  \hspace{1cm}  	\>1    \\
\hspace{1cm}$S_{ia}$ \hspace{1cm} \>1  \hspace{1cm}     \>1 \hspace{1cm}        \>1 \hspace{1cm}        \>1       
    
\end{tabbing}

\bigskip

\vspace{0.2cm}

\noindent with superpotential 

\vspace{0.2cm}

$W= \mu ZS + \lbrace Mq{\bar q} + Sp{\bar x} + q{\bar x}{\bar l} + B_1 p {\bar q} \rbrace + \mu S_{ia}(M_{ia}-{m^2 \over \mu} \delta_{ia})$. \hfill (3) 

\vspace{0.2cm}

\noindent Here $\mu$ denotes the introduced mass scale parameter which makes the introduced tree level terms in Eq.(3) have the correct mass dimension. The SU($n$) dynamics generates a scale $\Lambda_n$ which relates to $\Lambda_5$ by[1]

\vspace{0.2cm}

$\mu^{5+n} \sim \Lambda_5^{10-n} \Lambda_n^{2n-5}$. \hfill (4)

\vspace{0.2cm}

\noindent For $10 >  n \ge 3$, $\Lambda_n$ becomes strong while $\Lambda_5$ becomes weak and vice versa. For $n\geq 4$, the SU($n$) coupling is asymptotically free and the $SU(2)_1$ coupling is not. Let $\Lambda_{12}$ denote the $SU(2)_1$ scale. If the further assumption is made that $\Lambda_n \ll {m^2 \over \mu} < \mu \ll \Lambda_{12}$, we have 

\vspace{0.2cm}

$\Lambda_n \ll \mu \ll \Lambda_5$. \hfill (5)

\vspace{0.2cm}

\noindent It is clear from Eq.(5) that all S, Z, $S_{ia}$, $M_{ia}$ and $q_i$, $\bar q_i$ are heavy and can be integrated out. The resulting theories are the models of gauge SU($n$) $\times$ $SU(2)_1$ group with field content

\begin{tabbing}
\hspace{3cm}          \=SU($n$) \hspace{1cm} 		\= $SU(2)_1$ \hspace{1cm}  \\
\hspace{1cm} ${\bar x}$\hspace{1cm}       \>${\bar n}$ \hspace{1cm}  \>2  \hspace{1cm}      \\
\hspace{1cm}p \hspace{1cm}                \>n	 \hspace{1cm}	     \>1  \hspace{1cm}  	 \\
\hspace{1cm} ${q_{n+4}}$\hspace{1cm}      \> n	\hspace{1cm}	     \> 1 \hspace{1cm}	\\
\hspace{1cm}${\bar l}_i$\hspace{1cm}      \>1\hspace{1cm}	     \>2  \hspace{1cm}	 \\
\hspace{1cm}${\bar l}_{n+4}$ \hspace{1cm} \>1\hspace{1cm}	     \>2  \hspace{1cm}	 \\
\hspace{1cm}$B_{1i}$    \hspace{1cm}	  \>1 \hspace{1cm}	     \>1  \hspace{1cm}

\end{tabbing}

\vspace{0.2cm}

\noindent and superpotential

\vspace{0.2cm}

\hspace{1cm} $W={\bar x} q_{n+4} {\bar l_{n+4}} + {\mu \over m^2}{\bar l_i}B_{1i}p {\bar x}$ + {\large$ \lbrack {{\tilde {\Lambda}_n^{3n-2}} \over { \bar x \bar x p q_{n+4}}} \rbrack ^{1 \over {n-2}}$}, \hfill (6)

\vspace{0.2cm}

\noindent where $\Lambda_{12}$ denotes the $S(2)_1$ scale before integrating out heavy doublets and the last term of Eq.(6) is generated by the SU($n$) gauge dynamics. The new gauge scales $\tilde \Lambda_n$ and $\tilde \Lambda_{12}$ can be determined by the scale matching relations

\vspace{0.2cm}

\hspace{1cm} ${\tilde \Lambda_n}^{3n-2} = {\Lambda_n}^{2n-5} ({m^2 \over \mu})^{n+3}$ \hfill (7.1)

\vspace{0.1cm}

\hspace{1cm}$\Lambda_{12}^{n-3} = {\tilde \Lambda_{12}^{n-4}} \mu$. \hfill (7.2)

\vspace{0.2cm}

\noindent It can be shown that the tree level superpotential terms in Eq.(6) lift all classical flat directions and thus supersymmetry is broken by the generation of the non-perturbative superpotential.  The Yukawa coupling constant in Eq.(6) is in fact not small so that semi-classical calculation fails in estimating the vacuum energy. When $n$ is odd, we may also choose to add in the renormalizable terms $A{\bar Q_i}{\bar Q_j}$ to the original electric theory and integrate out the $SU(2)_1$ doublets ${\bar l_i}$. We are thus left with the resulting magnetic theories which are the various descriptions of the Intriligator-Thomas(IT) SU(n)$\times$SU(2) models[5] with extra singlets $B_1$.  The scale relations for the gauge dynamics again are given by

\vspace{0.2cm}

\hspace{1cm} ${\hat \Lambda_n}^{3n-2} = {\Lambda_n}^{2n-5} ({m^2 \over \mu})^{n+3}$ \hfill (8.1)

\vspace{0.1cm}

\hspace{1cm} ${\hat \Lambda_2}^{{11-n} \over 2} = \Lambda_{12}^{3-n} \mu^{{n+5} \over 2}$ \hfill (8.2)

\vspace{0.2cm}

\noindent where ${\hat \Lambda_2}$ and $\hat \Lambda_n$ denote the $SU(2)_1$ and SU($n$) scales of the IT models. For n=3, the dual model becomes the Affleck-Dine-Seiberg[9] $SU(3)\times SU(2)$ model with ${\hat \Lambda_2} > {\hat \Lambda_3}$. For n=5 and ${\hat \Lambda_2} \gg {\hat \Lambda_n}$, the resulting magnetic dual theory is the confined description of the IT model.  For n=5,7, or 9 and ${\hat \Lambda_n} \gg {\hat \Lambda_2}$, the magnetic theories are the electric descriptions of the IT models. For n=7, 9 and  ${\hat \Lambda_2} \gg {\hat \Lambda_n}$, the magnetic theories are the dual descriptions of the IT models.  For $n \ge 11$ the $SU(2)_1$ is not asymptotically free and the magnetic theories are free magnetic descriptions of other theories.  

\vspace{1cm}    
{\LARGE {\noindent 3. Case 2: $n=2$}}

\vspace{0.3cm}

Although Eq.(2) does not imply the classical breaking of supersymmetry, it is not a problem to have supersymmetry breaking magnetic models with gauge group SU(n)$\times$SU(2) when n=2. In fact, we can choose to add in tree level superpotential terms to the original electric theory which thus break supersymmetry classically:  

\vspace{0.2cm}

\hspace{0.8cm} $W_{tree}={S_{1ia}}(Q_i{\bar Q_a}-m_{ia})+{S_{2IJ} \over \mu}(A{\bar Q_I}{\bar Q_J}-h_{IJ})+{S_{3i} \over \mu}(AAQ_i-b_i)$. \hfill (9)

\vspace{0.2cm}

\noindent  Here $i$ runs from 1 to 5, $a$ runs from 1 to 6 and I, J run from 1 to 4. $S_{1ia}$, $S_{2IJ}$ and $S_{3i}$ are extra singlets and $m_{ia}$, $h_{IJ}$ and $b_i$ are parameters chosen to drive the theory to its strongly coupled regime. In general, we may have many different settings of mass parameters for the theory. For simplicity, we choose these parameters to preserve the global SP(2) symmetry:

\vspace{0.2cm}

$$m_{ia}=\cases{m^2 & for  (i,a)=(5,5), (4,6) \cr
&\cr
0 & otherwise. \cr}$$

$$h_{IJ}=\cases{m^3 & for (a,b)=(1,2), (3,4) \cr
&\cr
-m^3 & for (a,b)=(2,1), (4,3)    \cr
&\cr}$$

$$b_i=\cases{m^3 & for  i=1, \cr
&\cr
0 & otherwise. \cr}$$

\vspace{0.2 cm}

\noindent where $m$ is chosen to be much smaller than the original SU(5) gauge scale $\Lambda_5$. The above setting of parameters thus breaks supersymmetry classically even if we neglect the third term in Eq.(9). 

In terms of the field tensors of the deconfined electric theory, Eq.(9) can be rewritten as:

\vspace{0.2cm}

\hspace{1cm} $W_{tree}={\bar p}yz + S_{1ia}(Q_i{\bar Q_a}-m_{ia})+{S_{2IJ} \over {\Lambda_2 \mu }}(yy{\bar Q_I}{\bar Q_J}- \Lambda_2 h_{IJ})$

\vspace{0.1cm}

\hspace{2.2cm}$+{S_{3i} \over {\mu \Lambda_2^2}}(yyyyQ_i-{\Lambda_2^2}b_i)$. \hfill (10)

\vspace{0.2cm}
		 
\noindent Eq.(10) contains non-renormalizable terms which come from Eq.(9). To justify the deconfining method, it is implicitly assumed that the dynamical scale $\Lambda_2$ of the confining SU(2) group is much larger than $\Lambda_5$ in order for the SU(2) to confine. However, since the SU(5) gauge dynamics is in the free magnetic phase in the infrared, we can dualize it first and the dual theory can be found with field content listed below. 

\vspace{0.2cm}

\begin{tabbing}
\hspace{3cm}		  \=$SU(2)_d$ \hspace{1cm} 		\= $SU(2)_1$ \hspace{1cm} 	\= SU(5) \hspace{1cm} 	\=SU(6) \\
\hspace{1cm} ${\bar x}$\hspace{1cm}\>2 \hspace{1cm} 	   	\>2  \hspace{1cm}     	\>1   \hspace{1cm}      	\>1       \\
\hspace{1cm} ${\bar q}$\hspace{1cm}\>2   \hspace{1cm}    	\> 1   \hspace{1cm}    	\> ${\bar 5}$   \hspace{1cm}    		\>1 	 \\
\hspace{1cm} p\hspace{1cm} \>2	 \hspace{1cm}	\>1 \hspace{1cm}  	\> 1 \hspace{1cm}    		\>1	 \\
\hspace{1cm} q\hspace{1cm}\> 2	\hspace{1cm}		\> 1 \hspace{1cm}	\>1	\hspace{1cm}		\>$\bar 6$	\\
\hspace{1cm}${\bar l}$\hspace{1cm}\>1\hspace{1cm}	\>2\hspace{1cm}	\>1     \hspace{1cm}		\>6   \\
\hspace{1cm}M\hspace{1cm}	\>1\hspace{1cm}	\>1\hspace{1cm}	\>5    \hspace{1cm}	\>6 	\\
\hspace{1cm}$B_1$ \hspace{1cm}	\>1 \hspace{1cm}	\>1 \hspace{1cm}	\>5     \hspace{1cm}	\>1	\\
\hspace{1cm}S\hspace{1cm}	\>1 \hspace{1cm}	\>2 \hspace{1cm}	\>1   \hspace{1cm}      \>1    \\
\hspace{1cm}Z \hspace{1cm}	\>1 \hspace{1cm} 	\>2 \hspace{1cm} 	\>1  \hspace{1cm}  	\>1    \\  
\hspace{0.5cm}$S_{1ia}, S_{2IJ}, S_{3i}$  \>1 \hspace{1cm} 	\>1 \hspace{1cm} 	\>1  \hspace{1cm}  	\>1

\end{tabbing}

\vspace{0.2cm}

\noindent The magnetic superpotential must have extra tree level terms to eliminate unwanted states in the moduli space. 

\vspace{0.2cm}

\hspace{1cm}$W_{mag}=\mu ZS + \lbrace Mq{\bar q} + Sp{\bar x} + q{\bar x}{\bar l} + B_1 p {\bar q} \rbrace + \mu S_{1ia}(M_{ia}-{m_{ia} \over \mu} )$

\hspace{1.5cm} $+{\mu \over \Lambda_2} S_{2IJ} ({\bar l_I}{\bar l_J}-{{h_{IJ} \Lambda_2} \over \mu^2}) + \Lambda_2  S_{3i} (B_{1i} - {b_i \over {\mu \Lambda_2}})$. \hfill (11) 

\vspace{0.2cm}

\noindent Here $\mu$ again is a mass scale parameter. The last three terms in Eq.(11) are from the last three terms in Eq.(10) through operator mappings

\vspace{0.2cm}

\hspace{1cm}$Q_i {\bar Q_a} \sim  \mu M_{ia}$ \hfill (12.1)

\hspace{1cm}$yy{\bar Q_I}{\bar Q_J} \sim \mu^2 {\bar l_I}{\bar l_J}$ \hfill (12.2)

\hspace{1cm}$yyyyQ_i \sim \Lambda_2^3 {\bar p}Q_i  \sim  \Lambda_2^3 \mu B_{1i}$. \hfill (12.3)

\vspace{0.2cm}

\noindent Now let $\Lambda_{d2}$ and $\Lambda_{12}$ denote the gauge dynamics scales of the dual magnetic gauge groups $SU(2)_d$ and $S(2)_1$ respectively. Similar to the cases in section 2, the dualization and the deconfining method can only be applied when the following relations are satisfied

\vspace{0.2cm}

\hspace{1cm}$\mu^7 \Lambda_{d2} \sim \Lambda_5^8$ \hfill (13.1)

\hspace{1cm}$\Lambda_{12} \gg {m^2 \over \mu}$. \hfill (13.2)

\vspace{0.2cm}
\noindent If we further assume that the masses $\Lambda_2$ and $\mu$ are much larger than $\Lambda_{12}$ and smaller than $\Lambda_{d2}$, then heavy particles such as Z, S, $S_{1ia}$, $M_{ia}$, $S_{3i}$ and $B_{1i}$ can all be integrated out. The new scale $\tilde \Lambda_{12}$ of the $SU(2)_1$ after integrating out heavy sectors is found by scale matching:

\vspace{0.2cm}  

\hspace{1cm}$\Lambda_{12} \mu = {\tilde \Lambda_{12}}^2$. \hfill (14)

\vspace{0.2cm}

\noindent Together, Eqs.(13) and (14) imply

\hspace{1cm}$\Lambda_{d2} \gg \Lambda_5 > \mu \gg {\tilde \Lambda_{12}} \gg \Lambda_{12}$ \hfill (15.1)

\hspace{1cm}$\mu \gg m$. \hfill (15.2)

\vspace{0.2cm}

\noindent Eqs.(15.1,2) assure that the $SU(2)_1$ gauge dynamics is in the far infrared regime and can be dualized since there are now 8 doublets of $SU(2)_1$. The second magnetic dual theory thus has the matter content: 

\vspace{0.2cm}

\begin{tabbing}
\hspace{3cm}		          \=$SU(2)_d$ \hspace{1cm}  \= $SU(2)_2$ \hspace{1cm} 	\= SU(5) \hspace{1cm} 	 \=SU(6) \\
\hspace{1cm}    $x$\hspace{1cm}     \>2 \hspace{1cm} 	    \>2  \hspace{1cm}     	\>1   \hspace{1cm}       \>1       \\
\hspace{1cm} ${\bar q}$\hspace{1cm}\>2   \hspace{1cm}       \> 1   \hspace{1cm}    	\> ${\bar 5}$ \hspace{1cm}\>1 	 \\
\hspace{1cm} p\hspace{1cm}         \>2	 \hspace{1cm}	    \>1 \hspace{1cm}  	        \> 1 \hspace{1cm}    	  \>1	 \\
\hspace{1cm} q\hspace{1cm}         \> 2	\hspace{1cm}        \> 1 \hspace{1cm}	        \>1	\hspace{1cm}	  \>$\bar 6$	\\
\hspace{1cm} $l$\hspace{1cm}        \>1\hspace{1cm}           \>2\hspace{1cm}	         \>1     \hspace{1cm}	  \>$\bar 6$   \\

\hspace{1cm}$\bar a$ \hspace{1cm} \>1 \hspace{1cm}	    \>1 \hspace{1cm}	         \>1     \hspace{1cm}	  \>1	\\
\hspace{1cm}$\bar N$ \hspace{1cm} \>2 \hspace{1cm}	    \>1 \hspace{1cm}	         \>1     \hspace{1cm}	  \>6	\\ 
  \hspace{1cm}$H$ \hspace{1cm}	  \>1 \hspace{1cm}	    \>1 \hspace{1cm}	         \>1     \hspace{1cm}	 \>15	\\  
 \hspace{1cm}$S_{2IJ}$ \hspace{1cm}	  \>1 \hspace{1cm}	    \>1 \hspace{1cm}	         \>1     \hspace{1cm}	 \>1 \\  

\end{tabbing}

\vspace{0.4cm}

\noindent with more tree level terms added to superpotential

\vspace{0.2cm}

\hspace{1cm} $W=\lbrace {m^2 \over \mu}({q_5{\bar q_5}+q_6{\bar q_4}}) + {\tilde \mu}q{\bar N}  +  {m^3 \over{\mu \Lambda_2}}p {\bar q_1} \rbrace +  \lbrace {\bar a}xx + lx{\bar N}+Hll \rbrace$

\hspace{1.8cm} $+{{\mu {\tilde \mu}} \over \Lambda_2} S_{2IJ}(H_{IJ}-{{h_{IJ} \Lambda_2} \over {\mu^2 \tilde \mu}})$. \hfill (16)

\vspace{0.2cm}

\noindent Here ${\tilde \mu}$ is the mass scale introduced to relate the electric quark-antiquark bound states to the singlets in the magnetic theory. Let $\tilde \Lambda_{d2}$ and $\Lambda_{22}$ denote the scales of the new $SU(2)_d$ group and $SU(2)_2$ group respectively. The scale relation again is given by the previous dualization:

\vspace{0.2cm}

\hspace{1cm}$\Lambda_{22}^2 {\tilde \Lambda_{12}^2}={\tilde \mu^4}$ \hfill (17.1)

\hspace{1cm}${\tilde \Lambda_{d2}^4} \sim \mu^3 \Lambda_{d2}$. \hfill (17.2) 

\vspace{0.2cm}

\noindent It is noted that some Yukawa coupling terms in Eq.(11) become mass terms in Eq.(16). By assuming that all fields are suitable degrees of freedom and all massive terms that appear in Eq.(16) are heavy, 

\vspace{0.2cm}

\hspace{1cm}${{\mu {\tilde \mu}} \over \Lambda_2} > {{m^3 \Lambda_2} \over {\mu^2 {\tilde \mu}}} \gg \Lambda_{22}$ \hfill (18.1)

\hspace{1cm}${\tilde \mu} > {m^2 \over \mu},{m^3 \over {\mu \Lambda_2}}$ \hfill (18.2)

\vspace{0.2cm}

\noindent we are able to integrate them out and are left with the supersymmetric gauge $SU(2)_c \times SU(2)_3$ model with matter content 

\vspace{0.2cm}

\begin{tabbing}
\hspace{3cm}		             \=$SU(2)_c$ \hspace{1cm} 	\= $SU(2)_3$ 	 \\
\hspace{1cm} $x$    \hspace{1cm}     \>2 \hspace{1cm} 	   	\>2              \\
\hspace{1cm} ${\bar q_2}$\hspace{1cm}\>2   \hspace{1cm}    	\>1            \\
\hspace{1cm} ${\bar q_3}$\hspace{1cm}\>2   \hspace{1cm}    	\>1            \\
\hspace{1cm} ${\bar q_4}$\hspace{1cm}\>2   \hspace{1cm}    	\>1            \\
\hspace{1cm} ${\bar q_5}$\hspace{1cm}\>2   \hspace{1cm}    	\>1            \\
\hspace{1cm}$l_5$\hspace{1cm}        \>1\hspace{1cm}	        \>2              \\
\hspace{1cm}$l_6$\hspace{1cm}        \>1\hspace{1cm}	        \>2              \\
\hspace{0.7cm}$\bar a$, $H_{56}$ \hspace{1cm}    \>1 \hspace{1cm}	        \>1         	\\
		 
\end{tabbing}

\noindent and tree level superpotential

\vspace{0.2cm}
\hspace{1cm}$ W_{tree}={\bar a} x^2+ H_{56} l_5 l_6 + {m^2 \over{{\tilde \mu} \mu}}({\bar q_5} l_5 x + {\bar q_4} l_6 x)$. \hfill (19)

\vspace{0.2cm}

\noindent The scale relations of the $SU(2)_c \times SU(2)_3$ gauge model are given by

\vspace{0.2cm}

\hspace{1cm}$\Lambda_{c2}^3 = {({m^3 \over {\mu \Lambda_2}}) ^7 \over {\tilde \Lambda_{d2}^4}}$ \hfill (20.1) 

\hspace{1cm}$\Lambda_{32}^2 = {{m^3 \Lambda_2 \Lambda_{22}} \over {{\tilde \mu} \mu^2}}$ \hfill (20.2)

\vspace{0.2cm}

\noindent where $\Lambda_{c2}$ and $\Lambda_{32}$ denote the scales of the $SU(2)_c$ group and $SU(2)_3$ group respectively. The scale $\Lambda_{32}$ can be easily shown to be much larger than $\Lambda_{c2}$ by using Eq.(17.2), Eqs.(18.1,2) and Eqs.(20.1,2). Therefore, the $SU(2)_3$ group confines first and generates the quantum constraint on the moduli space. Supersymmetry is thus broken due to the generated quantum constraint.

We can make another interesting observation about the pseudo-flat direction. Classically we observed a pseudo-flat direction $\bar a$ which should be lifted by quantum corrections in Eq.(19). The Yukawa coupling constant associated with $ {\bar a} x^2$ is not a small number by Eq.(19). However, due to the large Yukawa coupling constant, we do not have suitable tools to quantitatively analyze the model and give a number to the quantum lift of the pseudo-flat direction.  On the other hand, the quantum lift of the pseudo-flat direction is expected to be much smaller than the gauge confining scale thus near to the origin. This expected answer can be seen if we make a naive assumption that the vacuum expectation value of the operator $Q_1Q_2Q_3Q_4Q_5$ can be no larger than the order of the magnitude of $m^5$.  If

\vspace{0.2cm}

\hspace{1cm}$\langle Q_1Q_2Q_3Q_4Q_5 \rangle \sim m^5$, \hfill (21)

\vspace{0.2cm}

\noindent the dual operator mappings then give the following relation

\vspace{0.2cm}

\hspace{1cm}$ {\bar a} \sim {1 \over {\tilde \mu}} x x \sim {\mu  \over {\Lambda_5^4 \tilde \mu}} Q_1Q_2Q_3Q_4Q_5$. \hfill(22)   

\vspace{0.2cm}

\noindent Combining equations (21) and (22), we find:

\vspace{0.2cm}

\hspace{1cm}$\lambda = 1$ \hfill (23.1)

\hspace{1cm}$<{\bar a}> \sim {{\mu m^5} \over {\Lambda_5^4 {\tilde \mu}}}$. \hfill (23.2)

\vspace{0.2cm}

\noindent Since the $SU(2)_3$ gauge group is the strong group, the theory can be viewed as the supersymmetric gauge SU(2) theory with four doublets. The quantum lift of this pseudo-flat direction has been argued by Shirman[10] and estimated to be of the order $\lambda {\bar a} \gg \Lambda_{32}$ in the case where $\lambda$ is small. In the case where $\lambda$ is not small, by dual operator mappings, we estimated the lift of the pseudo-flat direction $\bar a$ as follows
 
\vspace{0.2cm}

\hspace{1cm}${\mu^3 \over {m^3 \Lambda_2^2}} > {1 \over {\tilde \mu^2}}$ \hfill (24.1)

\hspace{1cm} $ ({{\lambda <{\bar a}>} \over \Lambda_{32}})^4 = ({{\mu^4 m^7  \over {{\tilde \mu} \Lambda_5^8 \Lambda_2 \Lambda_{22}}}})^2 \sim ({{\mu^4 m^7 {\tilde \Lambda_{12}}} \over {{\tilde \mu^3} \Lambda_2 \Lambda_5^5}} )^2$

\hspace{2.4cm}$< {{\mu^{17} m^5 {\tilde \Lambda_{12}^2}} \over {\Lambda_2^8 \Lambda_5^{16}}} < ({\mu^{17} \over {\Lambda_2 \Lambda_5^{16}}})({{m^5 {\tilde \Lambda_{12}^2}} \over \Lambda_2^7})$ 

\hspace{2.4cm}$\ll 1$ \hfill(24.2)

\vspace{0.2cm}

\noindent Eq.(24.2) is consistent with the expected smallness of the quantum lift of the pseudo-flat direction in the supersymmetric non-Abelian gauge theories. It is also noted that the smallness of the lift is quite general and that the changes in the Yukawa coupling constants in Eq.(9) will not alter the conclusion in Eq.(24.2).  For example, if we choose $W_{tree}$ of the original electric theory to be:

\vspace{0.2cm}

$W_{tree}={S_{1ia}}(Q_i{\bar Q_a}-m_{ia})+{S_{2IJ} \over \Lambda_2}(A{\bar Q_I}{\bar Q_J}-h_{IJ})+{S_{3i} \over \Lambda_2}(AAQ_i-b_i)$, \hfill (25)

\vspace{0.2cm}

\noindent we still reach the same conclusion on the smallness of the pseudo-flat direction lifting because of the constraints imposed by the dualization and the integrating out procedures. Although we do not prove the validity of the assumption in Eq.(21), we do see a small ratio of the pseudo-flat direction to the strong gauge scale by connecting the electric baryon operator to the magnetic singlet multiplet. 

\vspace{1.0cm}

\noindent {\Large \bf 4. Conclusion}

\vspace{0.8cm}

It is known that some dynamically supersymmetry breaking models can be interrelated through confining or duality in various limits[5]. In this paper we showed that the supersymmetric gauge $SU(n) \times SU(2)$ theories with dynamically supersymmetry breaking can be interrelated to the supersymmetric chiral gauge SU(5) theories which have an antisymmetric tensor, $n+3$ fundamentals, $n+4$ antifundamentals and gauge singlets as matter content when $n \ge 2$. In all the models been investigated, we added in tree level superpotential which drove the original theories to their strongly coupled regime and justified the usage of Seiberg's duality. The added superpotential terms are chosen to break supersymmetry at the classical level and thus the breaking of supersymmetry in the quantum level is also expected. 

For $n=2$, the second magnetic dual theory has the product gauge group $SU(2)_c \times SU(2)_3$ where the $SU(2)_3$ group is in its confining phase in the low energy regime. In the limit that $SU(2)_3$ is much stronger than the $SU(2)_c$ dynamics, the pseudo-flat direction $ \langle {\bar a} \rangle$ is observed and expected to be lift by quantum corrections to the K\"{a}hler potential. Based on the simple assumption in Eq.(21), we estimated the lifting of the pseudo-flat direction $\lambda {\bar a}$ to be much smaller than the strong confining scale $\Lambda_{32}$. This estimation is consistent with physical intuition. 

When $n \ge 3$, we have the supersymmetric gauge $SU(n) \times SU(2)$ models which can be interrelated to the original electric chiral SU(5) models. Supersymmetry is also broken dynamically in these magnetic dual models. For n being odd, the resulting $SU(n) \times SU(2)$ models are various descriptions of the Intriligator-Thomas $SU(n) \times SU(2)$ models. 

In general, we can follow the same thinking of lines as those of this paper and interrelate DSB models with $SU(n) \times Sp({{M-3} \over 2})$ gauge groups  or $SU(n) \times Sp(n-1)$ gauge groups to those supersymmetric chiral SU(M) gauge models with an antisymmetric tensor, $n+3$ fundamentals and $n+4$ antifundamentals. Since the dualization procedure can be iterated k times to yield dual descriptions with gauge group $SU(n) \times \prod_{i=0}^k Sp(n-1)$, we may also expect to interrelate supersymmetry breaking models with such product gauge groups to the chiral SU(M) models with an antisymmetric tensor and $n+3$ fundamentals.    

\bigskip

\centerline { {\large \bf  Acknowledgments}}

\vspace{0.4cm}
We would like to thank Michael E. Peskin for useful discussions and Yuri Shirman for conversation on pseudo-flat directions. We also thank Mandeep Gill for comments on the manuscript.  The work was supported under the DOE contract DE-AC03-76SF00515. 

\bigskip 

\noindent {\Large \bf References}

\vspace{1cm}

\noindent [1] N. Seiberg,  {\em Nucl. Phys.} {\bf B435}(1995)129. 

\noindent \hspace{0.5cm} L. Intriligator, N. Seiberg, {\em Nucl. Phys.} {\bf B444}(1995)125.

\noindent \hspace{0.5cm} K. Intriligator, P. Pouliot, {\em Phys. Lett.} {\bf B353}(1995)471

\noindent \hspace{0.5cm} K. Intriligator,  {\em Nucl. Phys.} {\bf B448}(1995)187

\noindent \hspace{0.5cm} K. Intriligator, R.G. Leigh, M.J. Strassler,  {\em Nucl. Phys.} {\bf B456}(1995)567

\noindent \hspace{0.5cm} John Brodie, {\em Nucl. Phys.} {\bf B478}(1996)123.

\noindent \hspace{0.5cm} John H. Brodie, Matthew J. Strassler,{\bf hepth/9611197}

\vspace{0.2cm}
\noindent [2] M. Berkooz, {\em Nucl. Phys.} {\bf B452}(1995)513.

\noindent \hspace{0.5cm} P. Pouliot, {\em Phys. Lett. } {\bf B367}(1996)151.

\noindent \hspace{0.5cm} Tadakatsu Sakai, {\bf hep-th/9701155}. 
 
\vspace{0.2cm}
\noindent [3] K. Intriligator, Scott Thomas, {\bf hepth/9608046}.

\noindent \hspace{0.5cm} P. Pouliot, {\em Phys. Lett. } {\bf B367}(1996)151.

\vspace{0.2cm}
\noindent [4] K. Intriligator, N. Seiberg, {\em Phys. Lett.} {\bf B342}(1995)152;

\noindent \hspace{0.5cm} M. Dine, A. E. Nelson, Y. Nir, Y. Shirman, {\em Phys. Rev.}{\bf D53}(1996)2658. 

\noindent \hspace{0.5cm} K. Intriligator, S. Thomas, {\em Nucl. Phys.}{\bf B473}(1996)121.

\noindent \hspace{0.5cm} E. Poppitz, Y. Shadmi, S.P. Trivedi, {\em Phys. Lett.}{\bf B388}(1996)561.

\noindent \hspace{0.5cm} C. Csaki, L. Randall and W. Skiba, {\em Nucl. Phys.}{\bf B479}(1996)65.

\noindent \hspace{0.5cm} Chih-Lung Chou, {\em Phys. Lett.}{\bf 391}(1997)329. 

\vspace{0.2cm}
\noindent [5] K. Intriligator and S. Thomas, {\bf hepth/9608046}.

\vspace{0.2cm}
\noindent [6] N. Kitazawa and N. Okada, {\bf hepth/9701370}.

\vspace{0.2cm}
\noindent [7] S. Coleman and E. Weinberg, {\em Phys. Rev.}{\bf D7}(1973)1888.

\vspace{0.2cm}
\noindent [8] M. Berkooz, {\em Nucl. Phys.} {\bf B452}(1995)513. 
 
\vspace{0.2cm}
\noindent [9] I. Affleck, M. Dine and N. Seiberg, {\em Nucl. Phys.}{\bf B256}(1985)557.

\vspace{0.2cm}
\noindent[10] Yuri Shirman, {\em Phys. Lett.}{\bf B389}(1996)287.

\end{document}